\documentclass[preprint,preprintnumbers,amsmath,amssymb]{revtex4}

\usepackage{graphicx,times}

\begin{document}
\title{Understanding the low-frequency modes in disordered systems at
single-particle level}
\author{Peng Tan$^1$, Ning Xu$^2$, Andrew B. Schofield$^3$,  and Lei Xu$^1$}
\affiliation{$^1$Department of Physics, The Chinese University of Hong Kong, Hong Kong, China\\
$^2$Department of Physics,University of Science and Technology of China, Hefei, China\\
$^3$ The School of Physics and Astronomy, University of Edinburgh,
Edinburgh, UK}

\date{\today}

\maketitle

\textbf{Normal modes provide a fundamental basis for understanding
crucial properties of solids, such as the thermal conductivity, the
heat capacity and the sound propagation\cite{Phillips1981,
Angellscience1995, Frickscience1995}. While the normal modes are
excellently described by plane waves in crystals, they are far less
understood in disordered systems, due to the great difficulties in
characterizing the heterogeneous vibrational behaviors. Using
charged colloids with long-range repulsion, we successfully make
different disordered systems without any contact friction, whose
normal modes can be visualized at single-particle level. In these
systems, we directly tackle the long-time outstanding puzzle in
condensed matter physics: the microscopic origin of the
low-frequency modes in disordered systems. For the first time, we
experimentally clarify that the low-frequency modes are caused by
the collective resonance of relatively disordered particles (or soft
structures) coupled with long-wavelength transverse excitations,
settling this puzzle at single-particle level. Next to these
low-frequency modes in the density of states, we also observe a
plateau due to isostaticity, verifying the fundamental prediction of
jamming model \cite{SilbertPRL2005,WyartEPL2005,SilbertPRE2009}.
Moreover, we reveal the intrinsic correlation between the
low-frequency modes and the real dynamics, which may lead to a
universal mechanism for aging, melting and yielding.}

One important unsolved problem in condensed matter physics is to
understand the normal modes of disordered systems. Unlike crystals,
whose particles have identical surroundings and vibrate as plane
waves, the particles of disordered systems have different local
environment and oscillate distinctively. To fully understand their
vibrations, therefore, we must correlate individual particles'
motions with their local environment, most ideally at
single-particle level. Because of experimental difficulties,
however, such studies are rare, leaving important questions
unsolved. In particular, disordered systems have much more
low-frequency modes than crystals, which often form a broad boson
peak in the $D(\omega)/\omega^{d-1}$ spectrum. These low-frequency
modes are generally believed to cause the anomalous properties in
heat capacity, thermal conductivity and acoustic behaviors. The
origin and properties of these modes are still open to debate. A
number of models have been proposed. In the energy landscape model,
the excess low-frequency modes are explained as the signature from a
minima-dominated phase to a saddle-point-dominated
phase\cite{Grigeranature2003}. In the jamming model, the jamming
transition at zero temperature (point J) is found responsible: it
produces a plateau in $D(\omega)$ as the result of isostaticity, and
creates extra modes at low frequencies \cite{GrestPRL1982,
NagelNature1998, HernPRL2002, SilbertPRL2005, WyartEPL2005,
WyartPRE2005, SilbertPRE2009, NingPRL2009, NingEPL2010}. Detailed
numerical analysis of the low-frequency modes suggests their
possible origin as transverse phonons associated with defective soft
structures\cite{tanakanm2008}. Experimentally, Raman, X-ray, and
neutron scatterings are commonly used to measure the normal mode
spectrum. It is shown that the low-frequency modes have collective
character\cite{prl2004} and non-acoustic properties\cite{prl2008}.
Most recently, the experiments in colloidal systems begin to examine
the normal modes at single-particle level\cite{Zhangnature2009,
Bonnprl2009, Chenprl2010, Islamscience2010}. Despite extensive
studies, however, the microscopic origin of the low-frequency modes
has never been experimentally clarified, which is the main focus of
this study.

We suspend PMMA particles of diameter $a=2.0 \mu m$ (polydispersity
$<$ 3\%) in density and refractive-index matched solvent, between
two glass substrates (Methods). In the weakly polar solvent, the
particles and the walls carry the same type of charges and interact
with long-range repulsion \cite{YethirajNature2003,
LeunissenNature2005}. Since there is no particle-particle and
particle-wall contact, we completely eliminate the contact frictions
in our system. Using confocal microscopy, we track every particle's
motion with excellent spatial($15nm$) and temporal ($30 s^{-1}$)
resolutions. We study perfect 2D samples by one single particle
layer between two glass substrates. To quantitatively describe
individual particle's local environment, we directly measure each
particle's local orientational order parameter, $\Psi_{6i}$,
determined by the nearest neighbors' arrangement: $\Psi_{6i} = 1$
means perfect hexagonal arrangement around particle $i$ and
$\Psi_{6i} = 0$ indicates totally random arrangement(see Methods for
exact definition). We use different colors to represent the
different $\Psi_{6i}$ values in Fig.1a. Clearly our heterogeneous
samples contain both ordered and disordered particles, with
$\Psi_{6i}$ ranging from $0$ to $1$. We can also systematically vary
the overall amount of disorder: by increasing the number density, we
increase the particle interaction and achieve more ordered samples,
as shown by sample-A through sample-D in Fig.1a and Fig.1b.

For each sample, we track the motions of all particles for 2000
frames and extract the normal modes with covariance matrix
method\cite{Bonnprl2009,Chenprl2010,Islamscience2010}(Methods). This
method is valid for stable and metastable systems, in which no
particle should rearrange; we therefore check every frame to make
sure no particle rearranges for the entire process. Within the
measurement time, every particle moves around its well-defined
equilibrium position and the system is metastable; while for longer
time, thermal motions cause rearrangements and the system relaxes.
This relaxation limits our measurement time to 2000 frames. To
compare the behaviors of particles under different local
environment, in each sample we divide all particles into two groups:
the relatively disordered particles with $\Psi_{6i}$ less than the
system average ($\Psi_{6i}<\overline{\Psi}_{6i}$), and the
relatively ordered particles with $\Psi_{6i}$ larger than the system
average ($\Psi_{6i}>\overline{\Psi}_{6i}$). The comparison between
the two groups will clarify the special role of the disordered
particles in the low-frequency modes. Fig.1c shows the mean square
displacement(MSD) for the relatively disordered particles (labeled
as ``dis'') and the relatively ordered particles (labeled as
``ord'') in samples A and C. The disordered curves are higher than
the corresponding ordered ones, indicating the qualitatively larger
movements of the disordered particles. In addition, all curves are
below the dashed line of free diffusion, as the result of caging
effect.

From the particle tracking, we extract the normal modes at
single-particle level, over three orders of magnitude in frequency.
We plot the density of states, $D(\omega)$, in Fig.2a
\cite{normalizefootnote}. For each curve, we can clearly identify
two characteristic frequencies, $\omega^\dag$ and $\omega^\ddag$,
which divide $D(\omega)$ into three regimes: a rapidly rising
low-frequency regime ($\omega<\omega^\dag$), a medium-frequency
plateau ($\omega^\dag<\omega<\omega^\ddag$), and a high-frequency
regime ($\omega>\omega^\ddag$). For the first time, we
experimentally measure a plateau in $D(\omega)$, which is predicted
in the jamming model as the result of isostaticity
\cite{SilbertPRL2005, WyartEPL2005,WyartPRE2005}. This plateau did
not appear in previous measurements\cite{Bonnprl2009, Chenprl2010,
Islamscience2010}, possibly due to the effect of contact
frictions\cite{MartinPre2007}. From A to D, as system disorder
decreases, the number of the low-frequency modes reduces, the
plateau width between $\omega^\dag$ and $\omega^\ddag$ shrinks, and
the plateau height lowers. The diminishing of the plateau should
come from the reduced amount of isostatic particles. Fig.2b plots
the $D(\omega)/\omega^{d-1}$ spectrum and shows the boson peak in
samples C, D but not in A, B. This is again consistent with the
jamming model: the boson peak shifts to lower frequencies in less
compressed systems such as A, B\cite{SilbertPRL2005,
WyartEPL2005,WyartPRE2005}.

To illustrate the microscopic origin of the low-frequency modes, we
analyze the normal modes at single-particle level. Fig.3a shows
three typical modes in the three different regimes of sample A (see
Supplementary Information for other samples). The low-frequency mode
is quasi-localized with large-scale collective correlations,
consistent with the jamming model\cite{SilbertPRL2005,
SilbertPRE2009, WyartPRE2005, NingEPL2010}; while the other two
modes look rather random. Further inspection on the low-frequency
mode reveals that its large-amplitude vibrations mostly occur at the
disordered particle sites. To quantitatively verify this, we compute
the average squared amplitude of the relatively disordered
particles, $\overline{A^2}_{dis}$, and the relatively ordered
particle, $\overline{A^2}_{ord}$. Their ratio,
$\overline{A^2}_{dis}/\overline{A^2}_{ord}$, quantifies the relative
vibrational strength between the two groups. We plot this ratio at
all frequencies in Fig.3c. In all the samples,
$\overline{A^2}_{dis}/\overline{A^2}_{ord}$ is greater than one at
small $\omega$ but quickly drops to unity as $\omega$ increases.
This quantitatively proves that the disordered particles vibrate
stronger than the ordered ones at low frequencies. Moreover, the
transition frequency, where
$\overline{A^2}_{dis}/\overline{A^2}_{ord}$ approaches unity, can
separately define an $\omega^\dag$, which coincides exactly with the
one found in $D(\omega)$ (Fig.2a). These data strongly suggest that
all the low-frequency modes in the first regime
($\omega<\omega^\dag$) are caused by the large-amplitude resonance
of disordered particles, clarifying the microscopic origin of the
low-frequency modes.

While the polarization vectors in real space reveal the resonance,
their Fourier transform in $q$ space brings further insights. In
Fig.3b, we decompose the same modes of Fig.3a into transverse and
longitudinal Fourier components (see Methods), and show their $q$
space distribution within the first Brillouin zone
[$-\frac{\pi}{\sigma}<q_x,q_y<\frac{\pi}{\sigma}$] ($\sigma$ is the
typical distance between two neighboring particles). We focus on the
low-frequency mode in column one ($\omega<<\omega^\dag$): the
transverse component has much larger overall magnitude and dominates
the longitudinal one, verifying the transverse nature of the
low-frequency modes \cite{tanakanm2008}. Moreover, it mostly
concentrates on small $q$ values, indicating long-wavelength
transverse excitations. At the medium frequency
($\omega^\dag<\omega<\omega^\ddag$), the transverse component
becomes uniformly distributed. When the high frequency is reached
($\omega>>\omega^\ddag$), the magnitude at small $q$ decreases
sharply, making the center more ``black'' than the edge. The
longitudinal component keeps spreading throughout $q$ space as
$\omega$ increases.

To illustrate the overall evolution of the transverse and
longitudinal components, we plot their average magnitude at all
frequencies in Fig.3d. For low and medium frequencies, the
transverse curves dominate the longitudinal ones, consistent with
Fig.3b. More interestingly, each transverse curve also contains two
characteristic frequencies, $\omega^\dag$ and $\omega^\ddag$, that
coincide with the ones found in $D(\omega)$. The wonderful agreement
of $\omega^\dag$ and $\omega^\ddag$ determined by various approaches
is demonstrated in Fig.3e, which summarizes different properties of
the normal modes from different aspects. Combining all the
properties of the first regime ($\omega<\omega^\dag$), we now draw
the final conclusion on the low-frequency modes: \emph{they are
caused by the collective resonance of the relatively disordered
particles, coupled with long-wavelength transverse excitations}.
This conclusion is based on the direct measurements of every
particle's local environment and vibrational properties in real and
Fourier space, and may also apply to other disordered systems.

Besides the microscopic origin, we illustrate another interesting
property of the low-frequency modes: their ability to drive the real
dynamics\cite{Brito2007, CooperNatPhy2008}. By projecting the
long-time (2000 frames) displacement field onto the normal mode
basis (Methods)\cite{BritoSoftmatter2010}, we find the displacement
mainly falls on the low-frequency modes. Fig.4a and 4b compare the
displacement field and the superposition of ten most contributed
low-frequency modes in sample A, both vector fields look quite
similar. Their difference is shown in Fig.4c: the small residue
proves their excellent agreement unambiguously. Fig.4d shows the sum
of the projection probabilities, $|c_\omega|^2$, and quantitatively
confirms the major contributions from the low-frequency modes. The
fact that we can wonderfully map the complex displacements of
thousands of particles with only ten low-frequency modes is rather
astonishing. It demonstrates that although the real dynamics is
thermally driven, it is not random during a long-time interval;
instead it follows specific low-frequency modes. Similarly, we
suspect that aging, melting and yielding may also be initiated by
specific low-frequency modes, and speculate the excitation of these
modes as the universal mechanism for these general phenomena.

In this study, we experimentally illustrate the origin and
properties of the low-frequency modes in disordered systems. By
dividing all particles into different groups with respect to their
local environment ($\Psi_{6i}$), we explicitly demonstrate the
important role of the relatively disordered particles. We note that
$\Psi_{6i}$ is just one of the many parameters that can be used. For
example, grouping particles according to their fluctuations around
equilibrium positions (local Debye-Waller factor) can yield similar
results (Supplementary Information). Therefore, our grouping
approach can be generalized and applied to many other disordered
systems, thus providing a general method to analyze heterogeneous
systems \cite{Shintaninp2006, Tanakaprl08, Tanakanm2010}.

\noindent\textbf{Methods}

\noindent \textbf{Sample Preparation}. The PMMA particles are dyed
with NBD and grafted with Polyhydroxystearic acid (PHSA) polymers.
They are suspended in the mixture of Iododecane and Iodododecane
(volume ratio 1:4), forming density and refractive-index matched
systems. The PHSA polymers are slightly charged in the solvent,
yielding the long-range repulsive potential between particles. We
make 2D samples by trapping one layer of particles between two cover
glasses. The glass surfaces are also grafted with PHSA polymers, and
repel the particles from large distance.

\noindent\textbf{Local Orientational Order Parameter}. The local
orientational order parameter for particle $i$ is defined as:
$\Psi_{6i}=\frac{1}{n_{i}}\mid\Sigma_{m=1}^{n_{i}}e^{j 6\theta_{mi}}\mid$.
Here $n_i$ is the number of nearest neighbors of particle $i$, and
$\theta_{mi}$ is the angle between $\textbf{r}_{m}-\textbf{r}_{i}$
and the x axis. $\Psi_{6i}=1$ means the perfect hexagonal
arrangement of six nearest neighbors and $\Psi_{6i}=0$ means totally
random arrangement.

\noindent\textbf{Covariance Matrix Method}. From particle tracking,
we can construct the covariance matrix and solve for the normal
modes. The elements of the covariance matrix are defined as:
\begin{equation}
C_{i,j}=\langle(\textbf{u}_{i}(t)-\langle
\textbf{u}_{i}(t)\rangle)(\textbf{u}_{j}(t)-\langle
\textbf{u}_{j}(t)\rangle)\rangle
\end{equation}
where $\textbf{u}(t)$ is the particle displacement at time $t$, and
$i,j=1, . . . , 2N$ run over the N particles and their x and y
coordinates. $\langle \rangle$ means time average. Diagonalizing the
matrix yields eigenvectors and eigenvalues. Since the covariance
matrix and the dynamic matrix have the relationship:
$D_{i,j}=\frac{k_BT}{m}(C^{-1})_{i,j}$, the covariance matrix
eigenvectors give the polarization vectors of the normal modes, and
the covariance matrix eigenvalues,
$\lambda=\frac{k_BT/m}{\omega^2}$, yield the angular frequencies
$\omega$.

\noindent\textbf{Transverse and Longitudinal Components}. The
transverse and longitudinal components of the normal modes in
Fourier space are defined as:
\begin{equation}
f_{\omega,tr}(\textbf{q})=|\Sigma_{i}(\hat{\textbf{q}}\times
\textbf{e}_{i,\omega})\texttt{exp}(i\textbf{q}\cdot\textbf{r}_{i})|^{2}
-\mathrm{transverse}
\end{equation}
\begin{equation}
f_{\omega,lo}(\textbf{q})=|\Sigma_{i}(\hat{\textbf{q}}\cdot
\textbf{e}_{i,\omega})\texttt{exp}(i\textbf{q}\cdot\textbf{r}_{i})|^{2}-\mathrm{longitudinal}
\end{equation}
Here $\omega$ labels the mode. ${{\bf{e}}_{i}}$ and ${\bf{r}}_{i}$
are the polarization vector and the equilibrium position vector of
particle $i$. The summation is over all particles.

\noindent\textbf{Projection of the Real Dynamics onto the Normal
Modes}. We first obtain the displacement field during the time
window of $\tau$=2000 frames:
\begin{equation}
|\delta r(\tau)>=|\delta r(t_{0}+\tau)>-|\delta r(t_{0})>
\end{equation}
We then project it onto the normal mode $|\omega >$, with the
prefactor $c_{\omega}$ determined by:
\begin{equation}
c_{\omega}=\frac{<\omega|\delta r(\tau)>}{\sqrt{<\delta
r(\tau)|\delta r(\tau)>}}
\end{equation}
The prefactors satisfy $\Sigma |c_{\omega}|^{2}=1$ for a complete
set of modes. We find that for some specific low-frequency modes,
the $|c_{\omega}|^2$ values are particularly large. These modes can
therefore be regarded as the driving modes of the real dynamics.


\clearpage
\newpage
\begin{figure}
\includegraphics[width=6in]{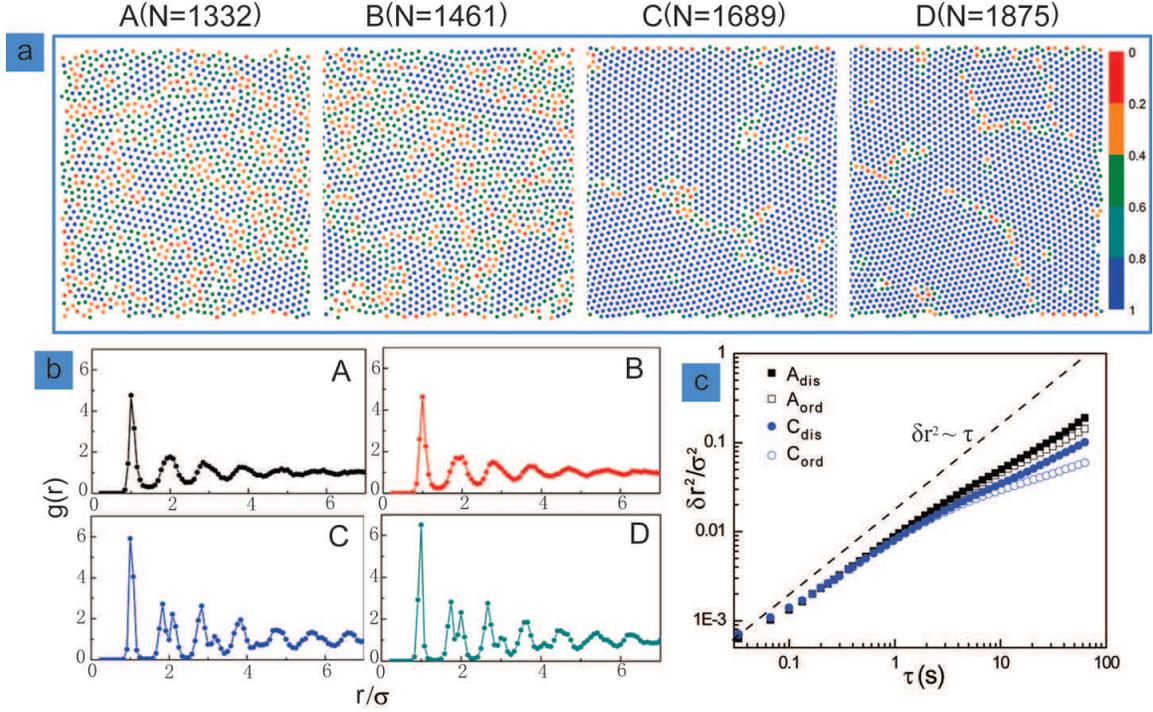}
\caption{\textbf{Structures and motions of four samples with
different amount of disorder}. \textbf{a}, the structures by the
particle equilibrium positions. The total dimensions are $145 \times
145\mu m^2$. Different colors correspond to different $\Psi_{6i}$
values. \textbf{b}, The pair correlation function, $g(r)$, of the
four samples. The distance $r$ is re-scaled by the position of the
first-peak, $\sigma$ (1.92$a$, 1.89$a$, 1.81$a$ and 1.73$a$ from A
to D). The peaks become more pronounced from A to D, indicating more
order in the sample. \textbf{c}, MSD curves for the relatively
disordered particles (solid symbols) and the relatively ordered
particles (open symbols) in A and C. Higher MSD curves of the
disordered particles demonstrate their larger motions. All curves
are below the dashed line of free diffusion, as the result of caging
effect.}\label{fig1}
\end{figure}

\clearpage
\newpage
\begin{figure}
\includegraphics[width=6in]{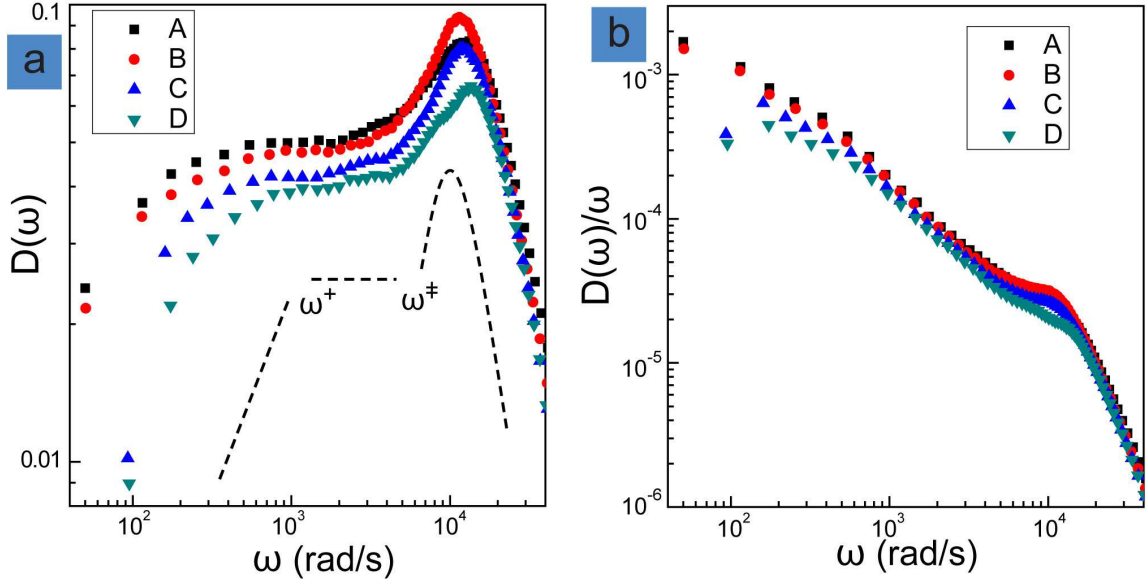}
\caption{\textbf{The distribution of the normal modes}. \textbf{a},
the density of states, $D(\omega)$, of the four samples. Each curve
can be divided into three regimes by two characteristic frequencies,
$\omega^\dag$ and $\omega^\ddag$, as illustrated by the schematics.
From A to D, as system disorder decreases, the number of the
low-frequency modes reduces, the plateau width between $\omega^\dag$
and $\omega^\ddag$ shrinks, and the plateau height lowers.
\textbf{b}, the $D(\omega)/\omega^{d-1}$ spectrum. The Boson peak
appears in C, D but shifts to lower frequencies in A, B. In both
\textbf{a} and \textbf{b}, the highest frequency plotted is
determined by our spatial resolution, $15 nm$.}\label{fig2}.
\end{figure}

\clearpage
\newpage
\begin{figure}
\includegraphics[width= 4.5 in]{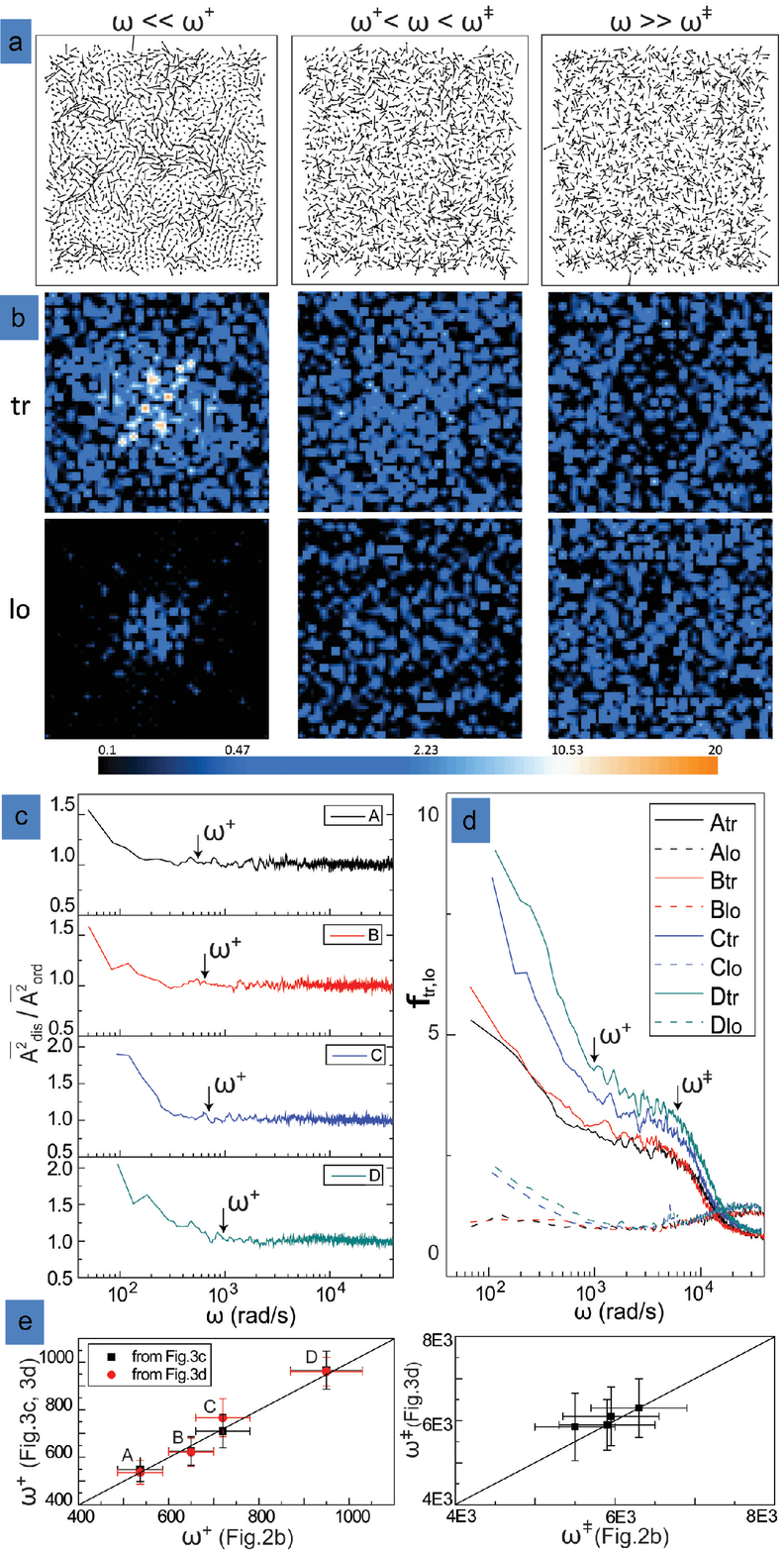}
\caption{}\label{fig3}
\end{figure}

\clearpage
\newpage
\noindent FIG. 3: \textbf{Vibration distribution in real and Fourier
space}. \textbf{a}, The polarization vectors of three typical modes
in sample A. The three typical modes illustrate the properties of
the three regimes. Clearly the low-frequency mode
($\omega<<\omega^\dag$) has large-scale correlations, while the
other two look rather random. \textbf{b}, The transverse and
longitudinal components of the same modes in Fourier space. Both
components are shown in the first Brillouin zone,
[$-\frac{\pi}{\sigma}<q_x,q_y<\frac{\pi}{\sigma}$]. As $\omega$
increases, the transverse component at small $q$ varies from very
large to very small values; while the longitudinal component keeps
spreading throughout $q$ space. \textbf{c}, the ratio of the mean
squared vibration amplitude between the disordered particles and
ordered particles, $\overline{A^2}_{dis}/\overline{A^2}_{ord}$. For
small $\omega$, the ratio is larger than one, indicating stronger
vibrations of the disordered particles. The transition frequency
where it approaches unity defines $\omega^\dag$ separately.
\textbf{d}, the average magnitude of transverse and longitudinal
components at all frequencies. The average is taken over the small
$q$ area at the center,
[$-\frac{\pi}{2\sigma}<q_x,q_y<\frac{\pi}{2\sigma}$]. The transverse
curves (solid) dominate the longitudinal ones (dashed) mostly. We
can again divide each transverse curve into three regimes, with two
characteristic frequencies, $\omega^\dag$ and $\omega^\ddag$.
\textbf{e}, the comparison of $\omega^\dag$ and $\omega^\ddag$
determined by different approaches. They excellently collapse onto
the straight line of $y=x$, confirming the robustness of the three
regimes determined differently.

\newpage
\begin{figure}
\includegraphics[width=5 in]{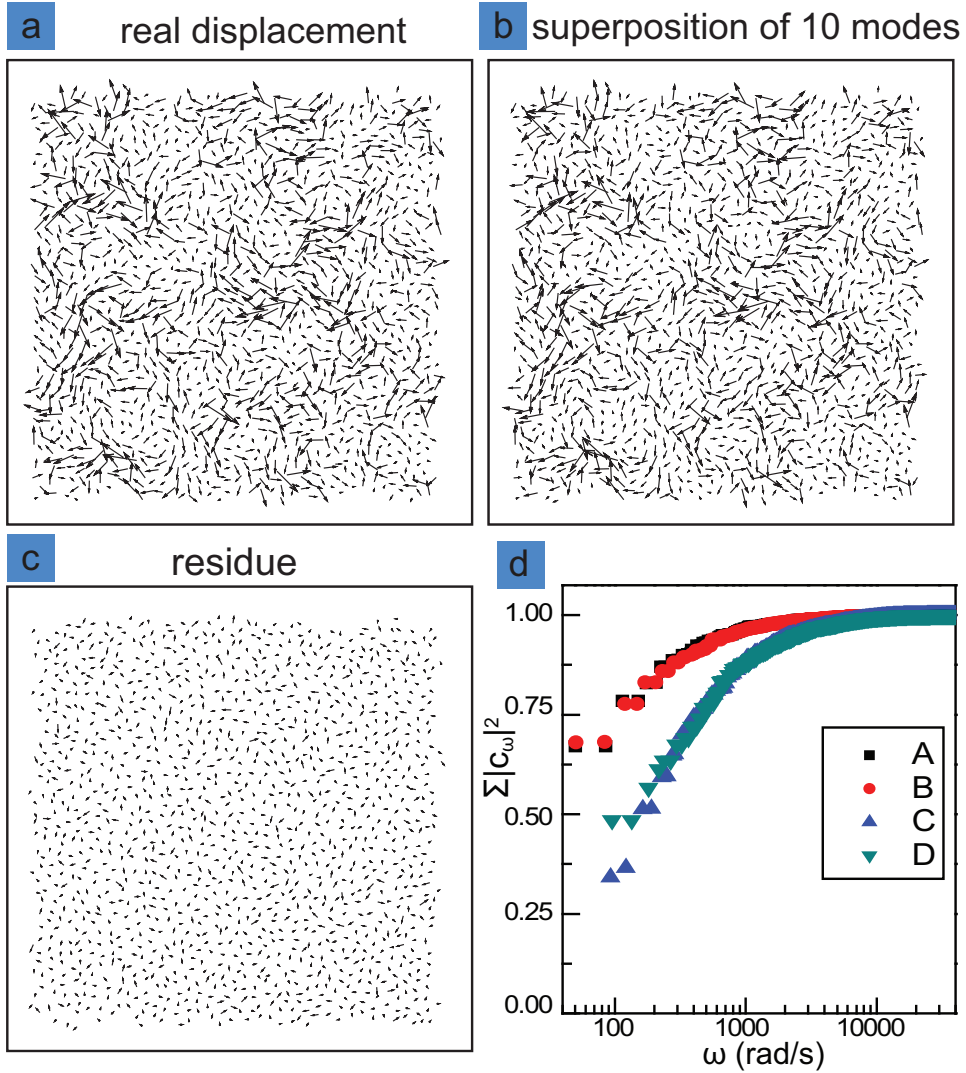}
\caption{\textbf{The comparison between the real dynamics and the
superposition of low-frequency modes}. \textbf{a}, the displacement
field during 2000 frames in sample A (enlarged 2.5 times for better
illustration). \textbf{b}, the superposition of ten most contributed
low-frequency modes. \textbf{c}, the residue from
\textbf{a}-\textbf{b}. The small residue proves excellent agreement.
\textbf{d}, the sum of the projection probabilities, $|c_\omega|^2$,
for all the modes. Apparently the low-frequency modes make major
contributions.}\label{fig4}
\end{figure}

\end{document}